\documentclass[12pt,oneside,english]{amsart}
\usepackage[T1]{fontenc}
\usepackage[latin9]{inputenc}
\usepackage{geometry}
\usepackage{amsthm}
\usepackage{amssymb}
\usepackage{graphicx}
\usepackage{esint}

\makeatletter
\numberwithin{equation}{section}
\numberwithin{figure}{section}
\newcommand{\id}{\textrm{d}}
\@ifundefined{date}{}{\date{}}
\makeatother

\usepackage{babel}
\begin{document}

\title{Friction and noise for a probe in a nonequilibrium fluid}

\author{Christian Maes}

\address{Instituut voor Theoretische Fysica, KU Leuven, Belgium}

\email{christian.maes@fys.kuleuven.be}

\author{Stefano Steffenoni}

\address{Dipartimento di Fisica ed Astronomia --- INFN, Università di Padova,
Italy}

\email{stefano.steffenoni@studenti.unipd.it}
\begin{abstract}
We investigate the fluctuation dynamics of a probe around a deterministic motion induced by interactions with driven particles.  The latter constitute the nonequilibrium medium in which the probe is immersed and is modelled as overdamped Langevin particle dynamics driven by nonconservative forces.  The expansion that yields the friction and noise expressions for the reduced probe dynamics is based on linear response around
a time-dependent nonequilibrium condition of the medium. The
result contains an extension of the second fluctuation--dissipation
relation between friction and noise for probe motion in a nonequilibrium
fluid. 
\end{abstract}
\maketitle
\renewcommand{\baselinestretch}{1.5}
{\large
\section{\textbf{The problem}}

Whether the environment of a system is in equilibrium or is driven into a nonequilibrium condition is important for characterizing internal motions and reactions.  In other words the reduced system dynamics will reflect whether the environment is driven or not.  For example, the relation between noise and friction on the system need no longer be described via the standard second fluctuation--dissipation (or Einstein) relation, we cannot unambiguously speak about entropy fluxes into the environment in terms of heat as there would be no Clausius relation, and system fluctuations or noise level in general are not simply quantified by a reservoir temperature.  The present paper takes up the challenge of characterizing such an effective dynamics for a probe in contact with a nonequilibrium medium.
One should have in mind that the medium consists of active elements or driven particles themselves in contact with a big thermal equilibrium reservoir (like surrounding air or water). Combined, the medium and the heat bath make up the nonequilibrium fluid. The system is called a probe here, but it can in general also refer to a collective or more macroscopic coordinate of the nonequilibrium fluid. 
The medium (both in contact with the probe and with the heat bath) consists of many ($N$) particles which we  model via a Langevin
dynamics  of the overdamped form for positions $x_{t}^{j}$,
\begin{equation}
\dot{x}_{t}^{j}=F\left(x_{t}^{j}\right)-\lambda\partial_{j}U\left(x_{t}^{j},X_{t}\right)-\underset{i<j}{\sum}\partial_{j}\Phi\left(x_{t}^{j}-x_{t}^{i}\right)+\sqrt{\frac{2}{\beta}}\,\xi_{t}^{j}\label{eq:fluid equation}
\end{equation}
with $j=1,\ldots, N$ labeling the particles and subject to independent
standard white noises $\xi^j_t$. The $\beta$ denotes the inverse temperature
of the background equilibrium reservoir but the particles are
effectively driven by the nonconservative force $F$. For simplicity we have not introduced explicitly friction and mass parameters, keeping only $\lambda$ (coupling) and $\beta$ as parameters.  The particle
interaction is given through the potential $\Phi$ while the potential $U$ depends on the
position $X_{t}$ of the probe, thus representing the back-reaction
of the probe on each particle. Other and different types of interaction between probe and medium are possible, e.g. in terms of their center of mass coordinate with a mean field type potential $U(\sum_j x_t^j- X_t)$, here not discussed and inessential for the present level of discussion.  Keeping to \eqref{eq:fluid equation} the probe dynamics itself is
\begin{equation}
M\ddot{X}_{t}+K\left(X_{t},\dot{X}_{t}\right)=-\lambda\underset{j}{\sum}\partial_{X}U\left(x_{t}^{j},X_{t}\right)\label{eq:probe equation}
\end{equation}
where we indicate by $K\left(X_{t},\dot{X}_{t}\right)$ other aspects
of the probe motion in the fluid.  Its mass $M$  is obviously also an important parameter for separation of time-scales between probe and fluid particle motion.

\medskip{}

The equations (\ref{eq:fluid equation}) and (\ref{eq:probe equation}) starting at $t=0$
are the basic evolution equations representing the coupled dynamics
of probe and fluid particles. Nonequilibrium works directly on the
medium which is both open to a thermal reservoir and is itself the environment of the probe. The problem of
the present paper is to characterize the effective or reduced probe
dynamics obtained from ``integrating out'' the fluid degrees of
freedom under the usual assumptions of weak coupling ($\lambda$ small)
for a large environment ($N$ big) which is rapidly relaxing ($M$ large) to every
new nonequilibrium steady condition as determined by the instantaneous
probe position.\\
In what follows we assume that the probe trajectory $[X] = (X_s, 0\leq s\leq t)$ deviates only a little
at least for some time span from a deterministic reference trajectory
$\left[Y\right]=\left(Y_{s},\,\,0\leq s\leq t\right)$ with $Y_0=X_0$,
which 
is a solution of an averaged out (\ref{eq:probe equation}),
\begin{equation}\label{fastl}
M\ddot{Y}_{t}+K\left(Y_{t},\dot{Y}_{t}\right)=-\lambda\underset{j}{\sum}\left\langle\partial_{X}
U\left(x^{j},Y_{t}\right) \right\rangle^{Y_t}
\end{equation}
where in the right-hand side we integrate out the positions $x^j$
of the medium particles 
using their stationary distribution $\langle \cdot\rangle^{Y_t}$ for the dynamics (\ref{eq:fluid equation})
at fixed probe position, i.e., evolving with
\begin{equation}
\dot{x}_{s}^{j}=F\left(x_{s}^{j}\right)-\lambda\partial_{j}U\left(x_{s}^{j},Y_{t}\right)-\underset{i<j}{\sum}\partial_{j}\Phi\left(x_{s}^{j}-x_{s}^{i}\right)+\sqrt{\frac{2}{\beta}}\,\xi_{s}^{j},\quad s>0\label{newfluid equation}
\end{equation}
In other words, to find $[Y]$ we apply an infinite time-scale separation between medium and probe motion so that the medium relaxes to its nonequilibrium steady condition at each fixed probe position, \cite{vK,KO}.  The trajectory $[Y]$ is then the supposedly unique solution to the probe equation \eqref{fastl} for given initial conditions.  We use the example of Section \ref{sec:Example} to make that more explicit.\\

One should think of $\left[Y\right]$
as the typical probe motion. 
 The nonequilibrium aspect of the medium is picked up in the effective force in the right-hand side of \eqref{fastl} and $[Y]$ will most likely not be constant in time; see also \cite{carl,miz,sheaopp,holz,drien} for statistical forces from nonequilibrium.  That can
be imagined as caused by some rotational force $F$ that acts on the
medium particles (as illustrated in Section \ref{sec:Example}), or it can be the combined result of strong nonlinearities
in the interaction $\Phi$.\\ 
When immersing a probe in the fluid we expect to see also friction
and noise on top of the behavior summarized in \eqref{fastl}. To understand their origin we must study the response of the medium particles to the stimulus of $X_s$ fluctuating around trajectory $[Y]$. 
The reference dynamics for the medium is 
\begin{equation}
\dot{x}_{s}^{j}=F\left(x_{s}^{j}\right)-\lambda\partial_{j}U\left(x_{s}^{j},Y_{s}\right)-\underset{i<j}{\sum}\partial_{j}\Phi\left(x_{s}^{j}-x_{s}^{i}\right)+\sqrt{\frac{2}{\beta}}\,\xi_{s}^{j},\quad s>0\label{adifluid equation}
\end{equation}
where we imagine $Y_s$ to be slowly changing.  The $[Y]$ is here  therefore a quasi-static protocol for the medium particles which at each moment $s$ are distributed using the stationary expectation $\langle \cdot\rangle^{Y_s}$.  Deviations from the real probe trajectory $h_{s}:=X_{s}-Y_{s}$
are supposed small for some good amount of time and hence the real dynamics \eqref{eq:fluid equation} is a perturbation of \eqref{adifluid equation}.  We can however find the new distribution of the medium particles by applying linear response theory. It is there that friction appears as the averaged deviation giving the reaction (or indeed,
response) of the medium particles to small $h_{s}$. In other words, the nonequilibrium fluid back-reacts
to the probe motion and makes the friction.  Noise is created due to the effectively random
 effect of the medium particles on the probe. The relation
between the noise covariance and the friction kernel is no longer
that of the standard Einstein or second fluctuation-dissipation relation.
That was already shown in \cite{key-1} in the same context as the
present paper but for $Y_{t}\equiv Y_0$ fixed in time.
The present paper is thus an extension of \cite{key-1} for probe
motion around (time-dependent) behavior. That requires
also an extension of the presently existing results for linear response
around nonequilibria. Here we need response theory against a time-dependent
background and the next section will start it with more general background collected in Appendix \ref{sec:Linear-response-theory}.  In Section \ref{sec:Reduced-dynamics}
we describe the coupled dynamics between fluid and probe and we specify friction and noise.  Section \ref{sec:Example} makes things more explicit for driven diffusive particles interacting with a probe in a toroidal trap.

\section{\textbf{Linear response around time-dependent nonequilibria}}
\label{sec:resp}

It suffices here to consider a single overdamped Langevin dynamics $x_{s}\in\mathbb{R}^{d}$ (representing a single medium particle)
in the presence of rotational forces $F$, confining potential $V$ and with a potential $U$
that depends on a time-dependent and deterministic
protocol $Y_{s}\in\mathbb{R}^{d}$,
\begin{equation}
\dot{x}_{s}=F\left(x_{s}\right)-\nabla_x V(x_s) -\nabla_x U\left(x_{s},Y_{s}\right)+\sqrt{\frac{2}{\beta}}\,\xi_{s}\label{eq:system unperturbed}
\end{equation}
Note that for simplicity we have not considered adding mobilities
or position-dependent diffusion constants (and we also put $\lambda=1$ when compared with \eqref{eq:fluid equation} or with \eqref{adifluid equation}); we just have standard $d-$dimensional
white noise $\xi_{s}$ with prefactor including the temperature $\beta^{-1}$
of a surrounding reservoir in thermal equilibrium. We assume that the
system for time-independent or fixed $Y_s\equiv a$ reaches a steady condition which can be described by a smooth
 density $\rho^a\left(x\right)$
on $\mathbb{R}^{d}$. It obviously requires that the potential $V$ is
sufficiently confining, as by the presence of a harmonic trap or a box with periodic
boundary conditions.  If moreover the protocol is quasi-static with respect to the sufficiently  short relaxation time of the medium particle, its position is distributed at each time $s$ by $\rho^{Y_s}$.

\medskip{}

We now perturb for $s>0$ as 
\begin{equation}
\dot{x}_{s}=F\left(x_{s}\right)-\nabla_x U\left(x_{s},Y_{s}\right)+\nabla_x h_{s}\cdot g\left(x_{s},Y_{s}\right)+\sqrt{\frac{2}{\beta}}\,\xi_{s}\label{eq:system perturbed}
\end{equation}
with small but arbitrary time-dependent amplitude $h_{s}$ and perturbing
potential $g\left(x,Y_{s}\right)$.  Note that besides the $h_s$, the time-dependence of the perturbation is again
through the same protocol $Y_{s}$. We could of course imagine other perturbation schemes  but for the present application, it suffices to consider the perturbation $Y_s\rightarrow Y_s + h_s$ which makes $g(x,Y_s) = - \nabla_{Y} U(x,Y_s)\in\mathbb{R}^{d}$.\\
  We assume that we start
at time zero from $\rho^{Y_0}$. The linear response we need here is the
difference in expectation $\delta_{g}A:=\left\langle A\left(x_{t}\right)\right\rangle^{[X]}-\left\langle A\left(x\right)\right\rangle^{Y_t}$
for observable $A$ to first order in $h_{s}$, $s\in\left[0,t\right]$,
where the first expectation is for the perturbed dynamics (\ref{eq:system perturbed}) corresponding to protocol $X_t := Y_t+ h_t$
and the second expectation $\left\langle A\left(x\right)\right\rangle^{Y_t} = \left\langle A\left(x_t\right)\right\rangle^{[Y]}$  is in the quasi--steady condition following
(\ref{eq:system unperturbed}). We give the derivation in Appendix \ref{sec:Linear-response-theory} and  here
is the result: 
\[
\delta_{g}A=\int_{0}^{t}\id s\, h_{s}\cdot\,R_{gA}\left(s,t\right)+O\left(h^{2}\right)
\]
with response coefficient given in terms of connected correlation functions $\langle w;v\rangle = \langle w\,v\rangle - \langle w\rangle\,
\langle v\rangle$,  in the reference quasi-steady condition of the medium particle, 
\begin{eqnarray}
R_{gA}\left(s,t\right) &=&\frac{\beta}{2}\frac{\id}{\id s}\left\langle g\left(x_{s},Y_{s}\right);A\left(x_{t}\right)\right\rangle^{[Y]}-\frac{\beta}{2}\dot{Y}_{s}\cdot\left\langle \nabla_Y g\left(x_{s},Y_{s}\right);A\left(x_{t}\right)\right\rangle^{[Y]}\nonumber\\&&
-\frac{\beta}{2}\left\langle L_{s}g\left(x_{s},Y_{s}\right);A\left(x_{t}\right)\right\rangle^{[Y]}\label{eq:system response}
\end{eqnarray}
Here $L_{s}$ is the instantaneous backward generator working on a
function $g$ of $x$ as
\begin{equation}\label{ls}
L_{s}g\left(x\right):=\left[F\left(x\right)-\nabla_x U\left(x,Y_{s}\right)\right]\nabla_x g\left(x\right)+
\frac{1}{\beta}\Delta_x g\left(x\right)
\end{equation}
Note that time zero has a special meaning in \eqref{eq:system response} as marking the beginning of the perturbed evolution.
There is another response formula that we need, expressing the change
in steady condition when we move $Y_{s}\rightarrow Y_{s}+a$
to a new protocol which deviates slightly, $||a||\ll1$. That
corresponds to taking $h_{s}\equiv a$ in (\ref{eq:system perturbed})
but also allowing sufficient time to reach a new steady condition. As we can
only apply (\ref{eq:system response}) when the perturbed and the
original dynamics have the same initial condition, we have to move
the latter to very early times to allow also the dynamics with protocol
$Z_s:=Y_{s}+a$ to reach its steady condition. In other words, we obtain from
(\ref{eq:system response}) the difference between two steady expectations, $\left\langle A\left(x\right)\right\rangle^{Y_t+a}-\left\langle A\left(x\right)\right\rangle^{Y_t} = $
\begin{eqnarray}
\left\langle A\left(x_{t}\right)\right\rangle^{[Z]}-\left\langle A\left(x_{t}\right)\right\rangle^{[Y]} &=& \frac{\beta a}{2}\left\langle g\left(x_{t},Y_{t}\right);A\left(x_{t}\right)\right\rangle^{[Y]}\nonumber\\
&-&\frac{\beta a}{2}\int_{-\infty}^{t}\id s\,\dot{Y}_{s}\cdot\left\langle \nabla_Y g\left(x_{s},Y_{s}\right);A\left(x_{t}\right)
\right\rangle^{[Y]}
\nonumber\\
&&
-\frac{\beta a}{2}\int_{-\infty}^{t}\id s\,\left\langle L_{s}g\left(x_{s},Y_{s}\right);A\left(x_{t}\right)\right\rangle^{[Y]}+O\left(a^{2}\right)\label{eq:static response}
\end{eqnarray}
where we have used that covariances $\left\langle g\left(x_{s},Y_{s}\right);A\left(x_{t}\right)\right\rangle^{[Y]}\rightarrow 0$
tend to vanish sufficiently fast as $s\downarrow -\infty$.\\
Formul{\ae} \eqref{eq:system response}--\eqref{eq:static response} provide extensions of linear response theory around
time-dependent nonequilibria. In previous work, e.g. \cite{key-4,key-8,key-13} about nonequilibrium linear response,
while in the same spirit, the nonequilibrium dynamics was not explicitly
time-dependent. We need these new formul{\ae} (derived in the Appendix) however for the application described in this paper.  Note also that we could have taken observable $A= A_t$ explicitly time-dependent and not only a function of the state at time $t$; we will need to apply the formul{\ae} to that case.

\section{\textbf{\label{sec:Reduced-dynamics}Reduced dynamics}}

The purpose of this Section is to obtain the reduced dynamics for the
probe. The result will be presented in terms of a Langevin equation where noise and friction originate from the interaction of the probe with the nonequilibrium fluid environment.
We must therefore integrate out the particles $x_t^j$ from  \eqref{eq:probe equation}.

Denote by 
\[
A(x,X) :=  -\lambda\,\sum_j\partial_{X}U\left(x^{j},X\right) 
\]
the force of the medium on the probe.  (We continue to use one--dimensional notation for simplicity.) The probe evolution equation \eqref{eq:probe equation} can be written
as the sum
of a deterministic and a random contribution, 
\begin{equation}
M\ddot{X}_{t}+K\left(X_{t},\dot{X}_{t}\right)= \left\langle A\left(x_{t},X_{t}\right)\right\rangle ^{\left[X\right]}+\eta_{t}\label{eq:noise in probe equation}
\end{equation}
defining the noise
\begin{equation}
\eta_{t}:=A\left(x_{t},X_{t}\right)-\left\langle A\left(x_{t},X_{t}\right)\right\rangle^{\left[X\right]}\label{eq:noise}
\end{equation}
We come back later to the  nature of that noise.
The first term on the right-hand side of \eqref{eq:noise in probe equation} is an average with respect to the fluid dynamics \eqref{eq:fluid equation} at $t>0$.  To first order in $X_t-Y_t$, that can be considered as the perturbed dynamics
\begin{equation}\label{perdy}
\dot{x}_{t}^{j}=F\left(x_{t}^{j}\right)-
\underset{i<j}{\sum}\,\partial_{j}\Phi\left(x_{t}^{j}-x_{t}^{i}\right)-\lambda\partial_{j}U\left(x_{t}^{j},Y_{t}\right)+h_{t}\partial_{j}
A\left(x_{t},Y_{t}\right)+\sqrt{\frac{2}{\beta}}\,\xi_{t}^{j}
\end{equation}
Remember here that $[Y]=(Y_s, s>0)$ is the reference probe trajectory, and that the coupling between probe and medium starts at time zero where $Y_0=X_0$. As \eqref{perdy} is of the form \eqref{eq:system perturbed} we rewrite \eqref{eq:noise in probe equation} as
\begin{eqnarray}
M\ddot{X}_{t}+K\left(X_{t},\dot{X}_{t}\right) &=& \left\langle A\left(x_{t},X_{t}\right)\right\rangle ^{\left[X\right]}   - \left\langle A\left(x_{t},X_{t}\right)\right\rangle ^{\left[Y\right]} \nonumber\\&+& \left\langle A\left(x_{t},X_{t}\right)\right\rangle ^{\left[Y\right]} +\eta_{t}\label{eq:noise in probe equationbis}
\end{eqnarray}
and we apply  the linear response formula \eqref{eq:system response} to the first line of \eqref{eq:noise in probe equationbis}.  
 There is a slight catch with respect to Section \ref{sec:resp} in that the observable $A=A_t$ is now also explicitly time-dependent,
$A_t\left(x_{t}\right)=A\left(x_{t},X_{t}\right)$.
Nevertheless the result \eqref{eq:system response} remains valid and translates here into
\begin{equation}
\left\langle A_t\left(x_{t}\right)\right\rangle ^{\left[X\right]}-\left\langle A_t\left(x_{t}\right)\right\rangle ^{\left[Y\right]}=\int_{0}^{t}\id s\,\left(X_{s}-Y_{s}\right)\frac{\id}{\id s}\mathcal{V}_t\left(s\right)\label{eq:susceptivity}
\end{equation}
where 
\begin{eqnarray}
\mathcal{V}_t\left(s\right)&:=& \frac{\beta}{2}\left\langle A\left(x_{s},Y_{s}\right);A_t\left(x_{t}\right)\right\rangle^{\left[Y\right]}
\nonumber \\
&-&\frac{\beta}{2}\int_{-\infty}^{s}\id u\left[\left\langle \dot{Y}_{u}\frac{\partial A\left(x_{u},Y_{u}\right)}{\partial Y};A_t\left(x_{t}\right)\right\rangle^{\left[Y\right]}\right]\nonumber\\
&-& \frac{\beta}{2}\int_{-\infty}^{s}\id u\left[\left\langle L_{u}A\left(x_{u},Y_{u}\right);A_t\left(x_{t}\right)\right\rangle^{\left[Y\right]}\right]
\label{46}
\end{eqnarray}
As in \eqref{ls}, $L_{s}$ is the backward generator for the unperturbed dynamics,
\begin{eqnarray}
L_{s}A\left(x_{s},Y_{s}\right)&=&
\underset{j}{\sum}\left(F\left(x_{s}^{j}\right)-\underset{i<j}{\sum}\partial_{j}\Phi\left(x_{s}^{j}-x_{s}^{i}\right)-\lambda\partial_{j}U\left(x_{s}^{j},Y_{s}\right)\right)\partial_{j}A\left(x_{s},Y_{s}\right)
\nonumber \\
&+& \beta^{-1}
\underset{j}{\sum}\partial^2_{jj}A\left(x_{s},Y_{s}\right)
\end{eqnarray}
We integrate  \eqref{eq:susceptivity} by parts using $h_{0}=0$,
\begin{equation}
\left\langle A_t\left(x_{t}\right)\right\rangle ^{\left[X\right]}-\left\langle A_t\left(x_{t}\right)\right\rangle ^{\left[Y\right]}
=
\left(X_{t}-Y_{t}\right)\mathcal{V}_t\left(t\right)-\int_{0}^{t}ds\,\left(\dot{X}_{s}-\dot{Y}_{s}\right)\mathcal{V}_t\left(s\right)\label{eq:partial integration}
\end{equation}

\subsection{Statistical force}
To rewrite the first term in the right-hand side of \eqref{eq:partial integration} we consider a static perturbation as in (\ref{eq:static response})
with $h_{t}\equiv h$. Since $h$ is now constant, the perturbed protocol $\left[Y+h\right]= (Y_s+h, s>0)$
is still quasi-static and the difference in steady components is described in (\ref{eq:static response}):
\begin{eqnarray}
\left\langle A_t\left(x_{t}\right)\right\rangle^{\left[Y+h\right]} =&&
\left\langle A_t\left(x_{t}\right)\right\rangle^{\left[Y\right]}+\frac{\beta h}{2}\left\langle A\left(x_{t},Y_{t}\right);A_t\left(x_{t}\right)\right\rangle^{\left[Y\right]}
\nonumber\\
-&&\frac{\beta h}{2}\int_{-\infty}^{t}\id u\,\left\langle \mathcal{K}_{u}\left(x_{u},Y_{u}\right);A_t\left(x_{t}\right)\right\rangle^{\left[Y\right]}
\end{eqnarray}
where
\begin{equation}
\mathcal{K}_{u}\left(x_{u},Y_{u}\right) := \dot{Y}_{u}\frac{\partial A\left(x_{u},Y_{u}\right)}{\partial Y}+L_{u}A\left(x_{u},Y_{u}\right)
\end{equation}
Comparing with \eqref{46} we can thus write
\begin{equation}
\left\langle A_t\left(x_{t}\right)\right\rangle^{\left[X\right]} = \left\langle A_t\left(x_{t}\right)\right\rangle^{\left[Y+h_{t}\right]}-\int_{0}^{t}\id s\,\left(\dot{X}_{s}-\dot{Y}_{s}\right)\mathcal{V}_t\left(s\right)\label{eq:result for A}
\end{equation}
and the temporal boundary term of \eqref{eq:partial integration} has yielded the zero order effective force on the probe:
\begin{eqnarray}\label{sf}
\eqref{eq:noise in probe equation} &=& G(X_t) -\int_{0}^{t}\id s\,\left(\dot{X}_{s}-\dot{Y}_{s}\right)\mathcal{V}_t\left(s\right)\nonumber\\
G\left(X_{t}\right) &:=& \left\langle A\left(x_{t},X_{t}\right)\right\rangle^{\left[Y+h_{t}\right]} = \langle A(x,X_t)\rangle^{X_t}
\end{eqnarray}
where the statistical average corresponds to taking the average over the medium in the right-hand side of \eqref{eq:probe equation} for infinite time separation between probe and fluid. That is exactly the reference dynamics \eqref{fastl}. Note that it could very well be that $G$ is a conservative force, and still the resulting probe dynamics will show fluctuations that betray the nonequilibrium nature of the medium.  That is encoded in the relation between friction and noise as comes next.

\subsection{Friction term}

The friction appears from the integral term in \eqref{eq:partial integration} and \eqref{eq:result for A}, with friction kernel
\begin{eqnarray}
\gamma\left(t,s\right) &:=& \frac{\beta}{2}\left\langle A\left(x_{s},Y_{s}\right);A\left(x_{t},Y_{t}\right)\right\rangle^{\left[Y\right]}
-\frac{\beta}{2}\,\int_{-\infty}^{s}\id u\left\langle \dot{Y}_{u}\frac{\partial A\left(x_{u},Y_{u}\right)}{\partial Y};A\left(x_{t},Y_{t}\right)\right\rangle^{\left[Y\right]}
\nonumber\\
&&\: - \frac{\beta}{2}
\,\int_{-\infty}^{s}\id u\left\langle L_{u}A\left(x_{u},Y_{u}\right);A\left(x_{t},Y_t\right)\right\rangle^{\left[Y\right]}
\label{eq:friction}
\end{eqnarray}
where we have worked to order $\lambda^2$ (replacing there $X_t$ with $Y_t$). Starting from \eqref{eq:noise in probe equationbis} and using \eqref{eq:result for A} with \eqref{eq:friction} we arrive at the probe effective evolution equation
\begin{equation}\label{redd}
M\ddot{X}_{t}+K\left(X_{t},\dot{X}_{t}\right) = G\left(X_{t}\right) - \int_{0}^{t}\id s\,\left(\dot{X}_s-\dot{Y}_s\right)\,\gamma\left(t,s\right) +\eta_{t}
\end{equation}
where we still need to discuss the noise $\eta_t$.

\subsection{Noise}

The noise is introduced in (\ref{eq:noise}). Its average is $\left\langle \eta_{t}\right\rangle ^{\left[X\right]}=0$, and the two-time correlations are
\[
\left\langle \eta_{s}\eta_{t}\right\rangle ^{\left[X\right]}
=
\left\langle A\left(x_{s},X_{s}\right);A\left(x_{t},X_{t}\right)\right\rangle ^{\left[X_{t}\right]}
\]
Considering the same perturbative regime as above, using the weak coupling ($\lambda$ small) we relate
these average values to the ones made over the quasi-steady condition for protocol 
$\left[Y\right]$:
\begin{equation}
\left\langle \eta_{s}\eta_{t}\right\rangle =
\left\langle A\left(x_{s},Y_{s}\right);A\left(x_{t},Y_{t}\right)
\right\rangle^{\left[Y\right]}\label{eq:noise expandend}
\end{equation}
to significant order.
Note that the noise perceived by the probe need not be Gaussian (we have not required a linear coupling between probe and medium nor did we specify the role of the interactions in the medium), and we have not insisted on obtaining a Markov (memoryless) limit.  That would require a more detailed study of time-scales.

\subsection{Second fluctuation--dissipation relation}

If we insert (\ref{eq:noise expandend}) in (\ref{eq:friction}) we
obtain
\begin{eqnarray}\label{fdr}
\gamma\left(t,s\right) &=& \frac{\beta}{2}\left\langle \eta_{s}\eta_{t}\right\rangle -\frac{\beta}{2}\int_{-\infty}^{s}\id u\left\langle \dot{Y}_{u}\frac{\partial A\left(x_{u},Y_{u}\right)}{\partial Y};A\left(x_{t},Y_{t}\right)\right\rangle^{\left[Y\right]}
\nonumber\\
&- &\frac{\beta}{2}\,\int_{-\infty}^{s}\id u\left\langle L_{u}A\left(x_{u},Y_{u}\right);A\left(x_{t},Y_{t}\right)\right\rangle^{\left[Y\right]}
\end{eqnarray}
which is different from the second fluctuation--dissipation relation valid for
systems in contact with an equilibrium environment, \cite{key-1,key-3,key-10,key-11,key-12}.  The modification to the standard fluctuation--dissipation relation (cf. \cite{rig,key-4}) can be given in terms of 
\[
{\mathcal X}_{st} := \frac 1{2}\big[ 1 + \frac{\int_{-\infty}^{s}\id u\left\langle \dot{Y}_{u}\frac{\partial A\left(x_{u},Y_{u}\right)}{\partial Y}+L_{u}A\left(x_{u},Y_{u}\right);A\left(x_{t},Y_{t}\right)\right\rangle^{\left[Y\right]}}{\left\langle A\left(x_{s},Y_{s}\right);A\left(x_{t},Y_{t}\right)
\right\rangle^{\left[Y\right]}}\big]
\]
Both friction kernel and noise are just functions of $t-s$ and assuming sufficient decay in memory so that $ {\mathcal X}_{st} = {\mathcal X}\,\exp[-\kappa (t-s)]$ for some large $\kappa>0$, 
we could call $\beta_{\text{eff}}^{-1} := ((1-{\mathcal X})\beta)^{-1}$ an {\it effective} temperature, as it would restore the Einstein relation 
\[
\gamma(t,s) = \beta_{\text{eff}}\,\langle\eta_s\eta_t\rangle
\]
but that is not quite sufficient for a thermodynamic meaning.  Moreover it makes sense to emphasize instead the {\it difference} between the terms in \eqref{fdr} that make the noise.  Following previous work on nonequilibrium linear response \cite{key-5,key-6,neg} we speak about an entropic and a frenetic contribution; see also Appendix \ref{sec:Linear-response-theory}.  The entropic part is purely dissipative and is proportional to the noise correlation; the frenetic component takes into account the changes in dynamical activity due to the perturbation.  The latter refers here to the time-symmetric activity of the medium particles and how that changes by a change in the probe position.

\section{\textbf{\label{sec:Example}Example}}
To understand what we have in mind for the reference protocol $Y_{t}$ or for the $X_t$ in \eqref{eq:fluid equation},
we consider here a  driven diffusion  first for one medium particle $x_{t}\in S^{1}$ on the circle $S^1$ of unit length,
\begin{equation}\label{exq}
\dot{x}_{t}=E-V'\left(x_{t}\right) - \lambda U'\left(x_{t},Y_t\right)+\sqrt{\frac{2}{\beta}}\,\xi_{t}
\end{equation}
The potential $V$ is periodic in $x\rightarrow x+1$ and $E>0$ is a constant driving.  There is a coupling with the probe at position $Y_t$ through the bi-periodic potential $U(x,Y)$.   Again $\xi_t$ denotes standard white noise and $U'(x,Y_t) = \partial_x U(x,Y_t)$.  We assume that the probe moves very slowly with respect to the medium.  As a result the particle moves around the circle, reaching
at each time a steady angular velocity
\begin{equation}
J_\lambda(t) = \left\langle E- V'(x) - \lambda U'\left(x,Y_t\right) \right\rangle^{Y_t}\label{eq:current}
\end{equation}
with expectation under the quasi-stationary probability density $\rho^{Y_t}\left(x\right),\, x\in S^{1}$.
This current and probability density should be
interpreted as follows. One considers a great many of such identical
independent medium particles with positions $x_{t}^{j}$ suspended in a viscous fluid in the same toroidal
trap modeled by $S^1$ and then $\rho^{Y_t}$ gives the real particle density in
the steady regime at probe position $Y_t$. Similarly, the mass center over all particles actually moves
with angular velocity $J_\lambda(t)$ given in \eqref{eq:current}. 

\medskip{}

The probe motion itself would naturally also be overdamped and following \eqref{fastl} we put
as reference dynamics 
\begin{equation}\label{refp}
\dot{Y}_t = \lambda\,\left\langle  U'\left(x,Y_t\right) \right\rangle^{Y_t}
\end{equation}
where we have used the action--reaction principle, \cite{hay}.  Comparing with \eqref{eq:current} we have
\begin{eqnarray}\label{rotpr}
\dot{Y}_t &=& E - \left\langle V'\left(x\right) \right\rangle^{Y_t} - J_\lambda(t)\nonumber\\
&=& J_0 - J_\lambda(t) + \left\langle  V'\left(x\right) \right\rangle -  \left\langle  V'\left(x\right) \right\rangle^{Y_t} 
\end{eqnarray}
where $J_0$ is the stationary current and $\langle\cdot\rangle$ is the stationary expectation at zero coupling $\lambda =0$. The reference probe motion \eqref{refp} is thus characterized as follows: (naturally) there is no motion for $\lambda=0$; for zero medium driving $E=0$ we have $J_0 =0= J_\lambda(t)$ and there may be various stationary points $Y_t\equiv y$ for which $\left\langle  V'\left(x\right) \right\rangle =  \left\langle  V'\left(x\right) \right\rangle^{y}$; for $V=0$ we see that the probe will move around the circle at an angular speed $J_0 - J_\lambda(t)$ which is  less than the current $J_0$ of the free medium particles in case the latter are slowed down by the probe ($J_\lambda(t) < J_0$) --- obviously, when the coupling $U$ is  rotation-invariant we have $J_\lambda(t) = J_\lambda$ independent of time and the probe just rotates at fixed speed.\\ 

The probe will not strictly follow
the dynamics \eqref{rotpr}, as there will be fluctuations due to the corpuscular nature of the nonequilibrium fluid. In other words, the probe has a
position $X_{t}$ not exactly equal to $Y_{t}$,  and therefore
the more correct Langevin equation for the medium particles is
\begin{equation}
\dot{x}_{t}^{j}=E-V'\left(x_{t}^{j}\right)-\lambda\,\partial_{j}U\left(x_{t}^{j},X_{t}\right)+\sqrt{\frac{2}{\beta}}\,\xi_{t}^{j}\label{eq:perturbed example}
\end{equation}
where $X_{t}$ moves around $Y_t$.  If
$X_{t}-Y_{t}$ is small we can expand the potential $U$ around $X_{t}=Y_{t}$ with
(\ref{rotpr}) thus representing zero order. As  in the previous section we can use the linear response formul{\ae} (\ref{eq:system response})--\eqref{eq:static response} to study how the probe motion perturbs
the fluid dynamics. From there, via \eqref{eq:probe equation}, the effective dynamics of the probe motion would appear.  Specific expressions will of course depend on the choices of the potential $V$ and $U$, but in principle everything needed for friction and noise covariaince is computable from two-time correlation functions under the quasi-stationary dynamics \eqref{exq}.

\section{\textbf{\label{sec:Outlook}Summary and Outlook}}

The character of particle motion in nonequilibrium fluids is a subject of increasing interest and is indeed relevant for a great variety of physical contexts ranging from motion in stellar and atmospheric environments to motion on the cytoskeleton of living cells; see e.g. \cite{key-9,drien,key-14,miz}.\\
Our set-up has been as follows:
\begin{itemize}
\item Driven particles make up the medium, both in contact with a probe and with a thermal equilibrium reservoir.  Their dynamics is modeled via a Langevin dynamics satisfying local detailed balance.
 We assume a reference probe trajectory obtained from an infinite time scale separation between probe and medium.
\item A reduced dynamics for the probe is obtained, in the form of a generalized Langevin equation containing a statistical force, friction and a noise term.  The derivation of the friction kernel is based
on linear response for the back reaction of medium particles on the motion of the probe which is considered to be a small perturbation of the reference trajectory.
\end{itemize}

There remain many interesting physical and computational points to be studied in the future and to be confronted with controlled experimental realizations.\\
The reduced probe dynamics is a nonequilibrium one, not satisfying the second fluctuation--dissipation relation.  One can try to consider introducing an effective temperature in terms of 
the ratio between the frenetic and the entropic component of the friction, but it remains to be been how useful that is for the probe's fluctuation and diffusion behavior.  More generally, response theory for this probe motion has not been physically discussed in the literature so far; the point is that we have no condition here of local detailed balance as there is no calorimetric way to speak about entropy fluxes in the nonequilibrium environment.\\
Another point of interest is to study the frenetic contribution to the friction; it could very well dominate when the environment is sufficiently far away from equilibrium and for example giving rise to the possibility of negative friction (as an analogue to what can happen in the out-of-equilibrium version of the first fluctuation-dissipation theorem \cite{neg}).\\
But even the statistical force $G$ is largely unexplored for nonequilibrium environments; it can certainly contain a rotational contribution which is a less studied topic in the general theory of fluctuation induced or Casimir forces. The simplest example was presented in the previous section for motion in a toroidal trap but even there more computational effort must be employed to give precise characterizations of the probe motion.

\medskip

}

\subsubsection*{\textbf{\emph{Acknowledgment}}}
\noindent{\bf Acknowledgment:} We thank Marco Baiesi and Urna Basu for their advice and for useful discussions.
This work was financially supported by the Belgian Interuniversity
Attraction Pole P07/18 (Dygest), and from an Erasmus fellowship allowing
S.S. to finish his Master thesis at the Institute for Theoretical
Physics at the KU Leuven.

\medskip
{\large
\appendix

\section{\label{sec:Linear-response-theory}Linear response theory}

\subsection{Girsanov Formula}

Consider the
perturbed Langevin equation 
\[
\dot{x}_{t}=\nu_{t}\left(x_{t}\right)\left[F_{t}\left(x_{t}\right)-\nabla U\left(x_{t}\right)\right]+\nabla D_{t}\left(x_{t}\right)+h_{t}\nu_{t}\left(x_{t}\right)\nabla V_{t}\left(x_{t}\right)+\sqrt{2D_{t}\left(x_{t}\right)}\,\xi_{t}
\]
where $h_{t}$ (nonzero for $t>0$) is a small time-dependent parameter. The perturbed
backward generator is
\[
L_{t}^{h}=\nu_{t}\left(x_{t}\right)\left[F_{t}\left(x_{t}\right)-\nabla U\left(x_{t}\right)\right]\cdot\nabla+\nabla\left(D_{t}\left(x_{t}\right)\cdot\nabla\right)+h_{t}\nu_{t}\left(x_{t}\right)\nabla V_{t}\left(x_{t}\right)\cdot\nabla
\]
involving a modification of the potential to $U_{t}^{h}\left(x_{t}\right) = U\left(x_{t}\right)-h_{t}V_{t}\left(x_{t}\right)$. 
We have assumed that the mobility and the diffusion coefficient have not changed.  We can then calculate the density of the perturbed probability $\mathcal{P}_{x_{0}}^{h}$ on trajectories $\omega$ over a time-span $[0,t]$ with respect to the original one $\mathcal{P}_{x_{0}}$ both starting from $x_0$.  In other words, there is an excess action ${\mathcal A}$, with formally
\[
\mathcal{P}_{x_{0}}^{h}\left(\omega\right) = \mathcal{P}_{x_{0}}\left(\omega\right)\,e^{-\mathcal{A}\left(\omega\right)}
\]
and, with $h_t\nu_t = D_t\varphi_t$,
\begin{eqnarray}
-\mathcal{A}\left(\omega\right) &=& \frac{1}{2}\int_{0}^{t}dx_{s}\circ\varphi_{s}\nabla V_{s}\left(x_{s}\right)\nonumber\\ &-&\frac{1}{2}\int_{0}^{t}ds\,\varphi_{s}\left[\nu_{s}\left(x_{s}\right)\left(F_{s}\left(x_{s}\right)-\nabla U\left(x_{s}\right)\right)\right]\nabla V_{s}\left(x_{s}\right)\nonumber\\
&-&\frac{1}{2}\int_{0}^{t}ds\,\varphi_{s}\nabla\left(D_{s}\left(x_{s}\right)\nabla V_{s}\left(x_{s}\right)\right)+\mathcal{O}\left(h_{s}^{2}\right)\label{eq:Girsanov Formula}
\end{eqnarray}
using Stratonovich stochastic integration in the first line. That is called a Girsanov formula and for diffusion processes very much resembles standard path-integration, see \cite{key-6}.  Note that in (\ref{eq:Girsanov Formula}) appears the backward generator
$L_s$.   We use the fundamental theorem of calculus to rewrite
\begin{eqnarray}
-\mathcal{A}\left(\omega\right) & = & \frac{1}{2}\left[\varphi_{t}V_{t}\left(x_{t}\right)-\varphi_{0}V_{0}\left(x_{0}\right)-\int_{0}^{t}\varphi_{s}\frac{\partial V_{s}\left(x_{s}\right)}{\partial s}ds -  \int_{0}^{t}\frac{d\varphi_{s}}{ds}V_{s}\left(x_{s}\right)ds\right]
\nonumber\\
&-&\frac{1}{2}\int_{0}^{t}ds\,\varphi_{s}\, L_{s}V_{s}\left(x_{s}\right)\label{eq:Girsanov form final}
\end{eqnarray}
Observe that the excess action $\mathcal{A}$ is made from two contributions: the first line is entropic
since it describes the excess entropy flux from the system to the environment due to the perturbation.
We mean excess because the system is out of equilibrium even without the perturbation;
there is an entropy production also in the unperturbed condition assured
by the external force $F_{t}\left(x_{t}\right)$. The second contribution describes the excess in dynamical activity. It takes
into account how much the system is inclined to change state.

\subsection{Response formula}

The perturbed
average values relate to the unperturbed ones via 
\begin{equation}
\delta\left\langle Q_{t}\left(x_{t}\right)\right\rangle _{\mu}^{h}=\left\langle Q_{t}\left(x_{t}\right)\right\rangle _{\mu}^{h}-\left\langle Q_{t}\left(x_{t}\right)\right\rangle _{\mu}\simeq-\left\langle \mathcal{A}_{t}\left(x_{t}\right)Q_{t}\left(x_{t}\right)\right\rangle _{\mu}\label{eq:susceptibility}
\end{equation}
where $\delta\left\langle Q_{t}\left(x_{t}\right)\right\rangle ^{h}$
is the generalized susceptibility related to the response by
\begin{equation}
\delta\left\langle Q_{t}\left(x_{t}\right)\right\rangle ^{h}\simeq\int_{0}^{t}\varphi_{s}\, R_{Q,V}\left(t,s\right)ds\label{eq:response}
\end{equation}
A simple inspection of \eqref{eq:Girsanov form final} gives
\begin{equation}
R_{Q,V}\left(t,s\right)=\frac{1}{2}\frac{d}{ds}\left\langle V_{s}\left(x_{s}\right)Q_{t}\left(x_{t}\right)\right\rangle _{\mu}-\frac{1}{2}\left\langle \frac{\partial V_{s}\left(x_{s}\right)}{\partial s}Q_{t}\left(x_{t}\right)\right\rangle _{\mu}-\frac{1}{2}\left\langle L_{s}V_{s}\left(x_{s}\right)Q_{t}\left(x_{t}\right)\right\rangle_{\mu}
\label{eq:response formula}
\end{equation}
The first two terms describe the entropic contribution. The last
term represents the frenetic part of the response; it depends on more detailed kinetics as here via the
mobility. This kind
of relation is a nonequilibrium extension of the (first) fluctuation--dissipation
theorem. 

}

\rule[0.5cm]{0.75\columnwidth}{1pt}

\end{document}